\documentclass{PoS}
\usepackage{cite}
\usepackage{amsmath}
\usepackage{amssymb}
\usepackage{epsfig}             
\usepackage{eufrak}
\usepackage{etex, xy}
\xyoption{all}

\newcommand{\ELI}{{\rm ELi}}

\newcommand*\pFqskip{8mu}
\catcode`,\active
\newcommand*\pFq{\begingroup
        \catcode`\,\active
        \def ,{\mskip\pFqskip\relax}%
        \dopFq
}
\catcode`\,12
\def\dopFq#1#2#3#4#5{%
        {}_{#1}F_{#2}\biggl[\genfrac..{0pt}{}{#3}{#4};#5\biggr]%
        \endgroup
}

\usepackage{tikz}
\usetikzlibrary{matrix}

\allowdisplaybreaks[4]

\title{
{\rm \footnotesize DESY 17-110, DO-TH 17/35, TTK-17-41, MSUHEP-17-022
}\\
Iterative and Iterative-Noniterative Integral Solutions in 3-Loop Massive QCD 
Calculations\thanks{This work was supported in part by the 
Austrian Science Fund (FWF) grant SFB F50 (F5006-N15, F5009-N15), FWF-Grant P 27229,
the European Commission through contract PITN-GA-2012-316704 ({HIGGSTOOLS}), 
and National Science Foundation under Grant No.~1719863.}}

\ShortTitle{Iterative and Iterative-Noniterative Integrals}

\author{J.~Ablinger, C.-S.~Radu, and C.~Schneider\\
Research Institute for Symbolic Computation (RISC),
                          Johannes Kepler University, Altenbergerstra\ss{}e 69,
                          A--4040, Linz, Austria}

\author{A.~ Behring, \\
Institut f\"ur Theoretische Teilchenphysik und Kosmologie, RWTH Aachen University,
Sommerfeldstr. 16, D-52074 Aachen, Germany}

\author{
J.~Bl\"umlein\footnote{Speaker}, A.~De Freitas\\
       Deutsches Elektronen-Synchrotron, DESY, Platanenallee 6, D-15738 Zeuthen, Germany}

\author{E.~Imamoglu, M.~van Hoeij\\
Department of Mathematics, Florida State University,
208 Love Building, 1017 Academic Way, Tallahassee, FL 32306-4510, USA}

\author{A.~von Manteuffel\\
Department of Physics and Astronomy, Michigan State University, East Lansing, MI 48824, USA
}

\author{C.G.~Raab\\
Institute for Algebra, Johannes Kepler University, Altenbergerstra\ss{}e 69, A-4040 Linz, Austria}

\abstract{Various of the single scale quantities in massless and massive QCD up to 3-loop order can be 
expressed by iterative integrals over certain classes of alphabets, from the harmonic polylogarithms 
to root-valued alphabets. Examples are the anomalous dimensions to 3-loop order, the massless Wilson coefficients 
and also different massive operator matrix elements. Starting at 3-loop order, however, also other letters appear 
in the case of massive operator matrix elements, the so called iterative non-iterative integrals, which are related 
to solutions based on complete elliptic integrals or any other special function with an integral representation
that is definite but not a Volterra-type integral. After outlining the formalism leading to iterative non-iterative 
integrals,we present examples for both of these cases with the 3-loop anomalous dimension $\gamma_{qg}^{(2)}$ and the 
structure of the principle solution in the iterative non-interative case of the 3-loop QCD corrections to the 
$\rho$-parameter.}

\FullConference{
RADCOR 2017 - 13th International Symposium on Radiative Corrections (Applications of Quantum Field Theory to 
Phenomenology) 24-29 September 2017,  St. Gilgen, Austria}

\makeatletter
\g@addto@macro\bfseries{\boldmath}
\makeatother
\begin{document}
\section{The Functional Structure of Feynman Integrals in the Single Scale Case}
\label{sec:1}

\vspace*{1mm}
\noindent
We will consider single scale quantities in massless and massive QCD up to 3-loop 
order. They are given by Feynman parameter integrals that depend only on one 
dimensionless parameter. Examples are the anomalous dimensions and the 
massless and massive Wilson coefficients, as well as the associated massive operator matrix elements (OMEs),
which can be represented by special functions of certain kinds. These are the iterative integrals up to the 2-loop 
case, which are given by
\begin{eqnarray}
H_{f_b,\vec{f_a}}(x) = \int_0^x dy f_b(y) H_{\vec{f_a}}(y),~~f_{c_i}(z) \in \mathfrak{A},~~H_\emptyset = 1,
\end{eqnarray}
with $\mathfrak{A}$ some alphabet of functions or distributions $f_{c_i}(z)$, or the related 
nested sums in 
Mellin $N$ space obtained by performing a Mellin transformation
\begin{eqnarray}
{\bf M}[H_{\vec{f_a}}(y)](N) = \int_0^1 dy y^{N-1}~H_{\vec{f_a}}(y).
\end{eqnarray}
The alphabet $\mathfrak{A}$ covers the harmonic polylogarithms \cite{Remiddi:1999ew}, Kummer-Poincar\'e iterated 
integrals \cite{Ablinger:2013cf}, cyclotomic harmonic polylogarithms \cite{Ablinger:2011te}, root-valued iterated 
integrals \cite{Ablinger:2014bra}, or generalizations thereof. In Mellin space the 
nested sums can be expressed in terms of the harmonic sums \cite{HSUM} and nested sums 
of Refs.~\cite{Ablinger:2013cf,Ablinger:2011te,Ablinger:2014bra} correspondingly.

All of these solutions have in common that their associated differential or 
difference equations are factorizable to first order.
They, or the difference equations obtained by using a formal power series Ansatz in solving the differential 
equations, can be completely solved using difference field and ring theory 
\cite{Karr:81,Schneider:01,Schneider:05a,Schneider:07d,Schneider:10b,Schneider:10c,Schneider:15a,
Schneider:08c,DFTheory}, as has been described in detail in Ref.~\cite{Ablinger:2015tua}. In the case of master 
integrals, this is possible for whatsoever basis. The corresponding alphabet $\mathfrak{A}$ is constructively found
in this way. These algorithms are implemented in the packages {\tt Sigma} \cite{SIG1,SIG2}, {\tt EvaluateMultiSums} 
and {\tt SumProduction} \cite{EMSSP}. In various places the package {\tt HarmonicSums} \cite{Ablinger:PhDThesis,
HARMONICSUMS,Ablinger:2011te,Ablinger:2013cf,Ablinger:2014bra} is used to operate on special functions of various
kind emerging throughout the calculation.

Beginning at 3-loop order a series of Feynman or master integrals obey differential 
equations which do not factorize at first order order anymore for the $O(\varepsilon^k)$ 
terms for $k \geq 0, k \in \mathbb{N}$, while for $k < 0$, first order factorization holds. 
This again can be proven by using the package {\tt Sigma}, which finds all the first 
order factors and will reduce the corresponding problem by leading to a remainder 
difference (differential) equation of order $ o > 1$.

The first new structures appear in the form of non-factorizing second order differential 
equations
\begin{eqnarray}
\left[\frac{d^2}{dx^2} + p(x) \frac{d}{dx} + q(x)\right] f(x)  =  N(x),
\label{eq:2}
\end{eqnarray}
containing usually more than three singularities and rational functions $p(x), q(x) \in 
\mathbb{Q}(x)$. Eq.~(\ref{eq:2}) can be a Heun \cite{HEUN} or a more general differential 
equation, i.e. more general than a Gau\ss{} differential equation. Yet one may find 
$_2F_1(a,b;c;z)$ homogeneous solutions, with rational parameters $a,b,c$ and $z(x) \in 
\mathbb{Q}(x)$. In even more general cases, one will have non-factorizable differential 
equations with rational coefficients of even higher order. Their solution is given in 
general by a combination of (known or yet unknown) higher transcendental functions. As we have outlined
in studying general single scale Feynman diagrams in \cite{Blumlein:2010zv}, the 
solutions can be obtained in the form of (multiple) Mellin-Barnes \cite{MB} integrals.
These integrals $F[r(y)]$ depend on one variable $y$, in such a way that the integral cannot be rewritten such 
that $y$ appears in the integration bounds only. Subsequently, further 
integrals shall be performed in the variable $y$ multiplying with other integrals of this type and iterative 
integrals. In this way,  non-iterative letters 
appear emerging in so-called iterative non-iterative integrals, 
\cite{Ablinger:2017bjx},
\begin{eqnarray}
\label{eq:ITNEW}
\mathbb{H}_{a_1,..., a_{m-1};\{ a_m; F_m(r(y_m))\},a_{m+1},...,a_q}(x) &=& \int_0^x 
dy_1
f_{a_1}(y_1) \int_0^{y_1} dy_2 ... \int_0^{y_{m-1}} dy_m f_{a_m}(y_m) F_m[r(y_m)]
\nonumber\\ &&
\times H_{a_{m+1},...,a_q}(y_{m+1}),
\end{eqnarray}
and $F[r(y)]$ is given by
\begin{eqnarray}
F[r(y)] = \int_0^1 dz g(z,r(y)),~~~r(y) \in \mathbb{Q}(y).
\end{eqnarray}

Writing the solutions in the case of second order differential 
equations as $_2F_1$ solutions has the advantage that a lot 
more is known about their functional properties, and all the Kummer relations and the 
contiguous relations can be applied. If and only if it is possible, after using these 
relations, in order to map the set $\{a,b;c\}$ into a certain finite set of triples, 
cf.~\cite{Ablinger:2017bjx}, there is a finite set of functional transformations 
of the $_2F_1$ solutions into complete elliptic integrals. 

In the $_2F_1$ solutions having appeared so far in single scale Feynman diagram 
calculations to higher order \cite{SABRY,TLSR1,Caffo:2002ch,Laporta:2004rb,
TLSR2,TLSR2a,Bailey:2008ib,Broadhurst:2008mx,TLSR3,BLOCH2,TLSR4,Adams:2015gva,
Adams:2015ydq,TLSR5,Adams:2014vja,Remiddi:2016gno,Adams:2016xah,TSLR6}, 
one could always find elliptic integral representations with either the 
appearance of only the complete elliptic integral of the first kind {\bf K} or of both 
the complete elliptic integrals of the first and second kind {\bf K} and {\bf E} in the single scale case.
The solutions of the second order differential equations read
\begin{eqnarray}
\label{eq:INHOM} 
\psi(x) &=&~~\psi_1^{(0)}(x) \left[C_1 - \int dx~\psi_2^{(0)}(x) n(x)\right]
 + \psi_2^{(0)}(x) \left[C_2 + \int dx~\psi_1^{(0)}(x) n(x) \right]~,
\label{eq:DI}
\end{eqnarray}
were $\psi_{1,2}^{(0)}(x)$ are the homogeneous solutions, $n(x)$ denotes the 
ratio of the inhomogeneity $N(x)$ and the Wronskian $W(x)$. $C_1$ and $C_2$ are 
constants depending on the physical problem.

One can now attempt to diagonalize (\ref{eq:DI}) by using modular functions, 
cf. e.g.~\cite{SERRE}, or in more special cases, modular forms. To do this one 
introduces the variable $q$
\begin{eqnarray}
q = \exp[-\pi {\rm \bf K}(1-z(x))/{\rm \bf K}(z(x))].
\end{eqnarray}
The original kinematic variable $x$ and all other building blocks of the 
inhomogeneous solution are then expressed in $q$.\footnote{In the cases studied so far the inhomogeneities
were always expressible in terms of harmonic polylogarithms. As the $q$ representation {\it depends on the 
process,} the representation of the harmonic polylogarithms have to be calculated for each case newly.}
To find 
the representation of 
$x=x(q)$ usually requires a higher order Legendre-Jacobi transformation, 
cf.~\cite{BB,Broadhurst:2008mx} through which $x(q)$ can be written as a
rational term of powers of Dedekind $\eta$ functions $\eta(k \tau),~~k \in 
\mathbb{N} \backslash \{0\}, \tau = \ln(q)/i$, \cite{DEDEKINDeta}, with
\begin{eqnarray}
\prod_{k=1}^m \eta^{l_k}(k \tau),~~~~l_k \in \mathbb{Z}.
\label{dedrat}
\end{eqnarray}
Each of the ratios (\ref{dedrat}), being modular functions under some 
congruence subgroup of $\Gamma(N)$ for some $N \in \mathbb{N}$, can now be written 
taking a factor $1/\eta^n(\tau)$ out for some $n \in \mathbb{N}$, such that the remainder term is a modular 
form. For all of these modular forms one may construct now a finite dimensional 
basis representation \cite{SERRE} through Lambert-Eisenstein series 
\cite{LAMBERT,EISENSTEIN} and products thereof \cite{COHEN}. They are of the form
\begin{eqnarray}
\label{eq:LE}
T_{m,n,l,a,b} &:=& \sum_{k=0}^\infty \frac{(mk + n)^{l - 1} q^{a (mk + n)}}{1 - q^{b 
(mk + n)}} \\ &=&
n^{l-1} q^{n(a-b)}
\text{Li}_0\left(q^{nb}\right)
+ q^{n(a-b)} \sum_{j=0}^{l-1} \binom{l-1}{j} m^j n^{l-1-j} 
\text{ELi}_{-j;0}\left(q^{m(a-b)};q^{nb};q^{mb}\right),
\nonumber
\end{eqnarray}
or special cases thereof.
Here Li$_0(x) = x/(1-x)$ and ELi$_{n,m}(x,y;q)$ denotes the 
elliptic polylogarithm \cite{Adams:2015ydq}
\begin{eqnarray}
\text{ELi}_{n,m}(x,y;q) = 
\sum_{k=1}^\infty
\sum_{l=1}^\infty 
\frac{x^k}{k^n} 
\frac{y^l}{l^m} q^{nm}. 
\label{eq:ELP}
\end{eqnarray}
Note that in Eq.~(\ref{eq:LE}) the parameters $x$ and $y$ of the elliptic 
polylogarithm (\ref{eq:ELP}), which are not supposed to depend on $q$, do depend on $q$.
One may synchronize arguments $q^m \rightarrow q, -q \rightarrow q$ within the notion
of elliptic polylogarithms, cf.~\cite{Ablinger:2017bjx}.
In the above sense, the elliptic polylogarithm provides a frame for the result, but not 
in the original sense.
Due to the multiplication relation of the elliptic polylogarithm all of these terms can 
be represented in the form of elliptic polylogarithms, which are formal power series 
in $q$. 

However, there are still the factors $1/\eta^k(\tau)$ for which the closed form 
representation of its formal power series in $q$ is yet unknown, unlike its infinite 
product representation \cite{COHEN}. This means that for the case that all $k=0$, the 
integral-relation of the elliptic polylogarithms will yield elliptic polylogarithms 
again. Up to terms of $\ln(q)$ the inhomogeneous solution (\ref{eq:DI}) can thus be 
diagonalized and be written mainly in terms of elliptic polylogarithms.
This is different in cases in which $k \neq 0$. Here the $q$-integrals will 
usually lead to higher transcendental functions in $q$, but will not be diagonalized in 
operating on $q$ series for which the expansion coefficients are known in closed form.

In the following we will illustrate the above formalisms applied to two important cases. One of them
is the 3-loop anomalous dimension, which can be expressed in $x$-space either in terms 
of iterative integrals, which are harmonic polylogarithms, or the nested harmonic 
sums in Mellin $N$-space. The 3-loop QCD corrections to the $\rho$-parameter, on the 
other hand, are an example in which elliptic integrals are present. At least in 
intermediary steps one needs modular functions, represented by modular forms as well as a 
pre-factor $1/\eta^k(\tau)$.  
\section{The 2-loop anomalous dimension $\gamma_{qg}^{(2)}$}
\label{sec:2}

\vspace*{1mm}
\noindent
In Ref.~\cite{Ablinger:2017tan} we calculated the 3-loop anomalous dimension $\gamma_{qg}^{(2)}(N)$ from first 
principles in a massive environment. It has been obtained from the $O(1/\varepsilon)$ term of the unrenormalized
3-loop OME $\hat{\hat{A}}_{Qg}^{(3)}$. Due to the polynomial dependence of the master integral on the dimensional
parameter $\varepsilon$, one would also encounter elliptic terms if the calculation of the master integrals 
would be carried out directly. The computational method we used was, however, the method of arbitrarily large fixed
moments \cite{Blumlein:2017dxp}. In this way, new higher transcendental functions of whatsoever complexity will 
map onto specific series of
rational numbers. These series are united to a series for $\hat{\hat{A}}_{Qg}^{(3)}$. Due to this, we obtain new series
in $N$ for the $O(1/\varepsilon)$ term, for all the individual color-$\zeta$ factors labeled by the corresponding
Casimir operators and potential multiple zeta values \cite{Blumlein:2009cf}. We have generated 2000 moments for 
$\hat{\hat{A}}_{Qg}^{(3)}$ using the formalism of Ref.~\cite{Blumlein:2017dxp} in
Ref.~\cite{Ablinger:2017tan}. Guessing methods 
\cite{GUESSHB,Blumlein:2009tj} now allow to find a difference equation for the different contributions to 
$\gamma_{qg}^{(2)}(N)$. It turns out that this equation, unlike the ones for the master integrals, which would be 
needed to higher powers in $\varepsilon$, factorizes all in first order. 
They can be solved by applying difference field and ring methods 
\cite{Karr:81,Schneider:01,Schneider:05a,Schneider:07d,Schneider:10b,Schneider:10c,Schneider:15a,
Schneider:08c,DFTheory} using the packages {\tt Sigma} \cite{SIG1,SIG2}, {\tt EvaluateMultiSums} and {\tt 
SumProduction} \cite{EMSSP}. Here the package {\tt HarmonicSums} \cite{Ablinger:PhDThesis,
HARMONICSUMS,Ablinger:2011te,Ablinger:2013cf,Ablinger:2014bra} is used to operate on special functions of various 
kind.

We obtain
\begin{eqnarray}
\gamma_{qg}^{(2)} &=&
        \textcolor{blue}{C_A N_F^2 T_F^2} \Biggl\{
                -
                \frac{128(5 N^2+8 N+10)}{9 N (N+1) (N+2)}  
S_{-2}
                -\frac{64 P_8}{9 N (N+1)^2 (N+2)^2} S_1^2
        \nonumber\\ &&
                -\frac{64 P_9}{9 N (N+1)^2 (N+2)^2} S_2 
                +\frac{64 P_{25}}{27 N (N+1)^3 (N+2)^3} S_1 
                +\frac{16 P_{34}}{27 (N-1) N^4 (N+1)^4 (N+2)^4} 
        \nonumber\\ &&
                +p_{qg}^{(0)}(N) \Biggl(
                        \frac{32}{9} S_1^3
                        -\frac{32}{3} S_1 S_2
                        +\frac{64}{9} S_3
                        +\frac{128}{3} S_{-3}
                        +\frac{128}{3} S_{2,1}
                \Biggr)
        \Biggr\}
        \nonumber\\ &&
        +\textcolor{blue}{C_F N_F^2 T_F^2}
        \Biggl\{
                        \frac{32(5 N^2+3 N+2) }{3 N^2 (N+1) (N+2)}  S_2
                        +\frac{32(10 N^3+13 N^2+29 N+6) }{9 N^2 (N+1) (N+2)}  S_1^2 
        \nonumber\\ &&
                        -\frac{32 
        P_{12}}{27 N^2 (N+1)^2 (N+2)} S_1
                        +\frac{4 P_{38}}{27 (N-1) N^5 (N+1)^5 (N+2)^4} 
        \nonumber\\ &&
        +p_{qg}^{(0)}(N) \Biggl(
                                -\frac{32}{9} S_1^3
                                -\frac{32}{3} S_1 S_2
                                +\frac{320}{9} S_3
                        \Biggr)
                \Biggr\}
        \nonumber\\ &&
                +\textcolor{blue}{C_A C_F N_F T_F} \Biggl\{
                        -128 \frac{ N^3-7 N^2-6 N+4 }{N^2 (N+1)^2 (N+2)} S_{-2,1}
                        +\frac{32 P_5}{N^2 (N+1)^2 (N+2)} S_{-3}
        \nonumber\\ &&            
            +\frac{16 P_{18}}{9 (N-1) N^2 (N+1)^2 (N+2)^2} S_1^3  
                        -\frac{16 P_{24}}{9 (N-1) N^2 (N+1)^2 (N+2)^2} S_3 
        \nonumber\\ &&   
                     -\frac{8 P_{27}}{9 (N-1) N^3 (N+1)^3 (N+2)^2} S_1^2
                        +\frac{8 P_{29}}{3 (N-1) N^3 (N+1)^3 (N+2)^3} S_2
        \nonumber\\ &&
                        +\frac{P_{37}}{27 (N-1) N^5 (N+1)^5 (N+2)^4} 
                        + p_{qg}^{(0)}(N) \Biggl[
                                \left(
                                        \frac{640}{3} S_3
                                        -384 S_{2,1}
                                \right) S_1
                                +\frac{32}{3} S_1^4
        \nonumber\\ &&
                                +160 S_1^2 S_2
                                -64 S_2^2
                                +\big(
                                        192 S_1^2
                                        +64 S_2
                                \big) S_{-2}
                                +96 S_{-2}^2
                                +224 S_{-4}
                                -64 S_{2,-2}
                                +64 S_{3,1}
        \nonumber\\ &&
                                +192 S_{2,1,1}
                                -256 S_{-2,1,1}
                                -192 S_1\zeta_3
                        \Biggr]
                        -\frac{192 P_{17}}{(N-1) N^2 (N+1)^2 (N+2)^2} \zeta_3
        \nonumber\\ &&             
           +\Biggl(
                                \frac{16 P_{16}}{3 (N-1) N^2 (N+1)^2 (N+2)^2} S_2
                                +\frac{16 P_{35}}{27 (N-1) N^4 (N+1)^4 (N+2)^4}
                        \Biggr) S_1
        \nonumber\\ &&
                        +\Biggl[
                                -\frac{32 P_{15}}{N^3 (N+1)^3 (N+2)}
                                +\frac{128 \big(N^3-13 N^2-14 N-2\big)}{N^2 (N+1)^2 (N+2)}  S_1
                        \Biggr] S_{-2}
        \nonumber\\ &&
                        +\frac{96 N (N+1) p_{qg}^{(0)}(N)^2 }{N-1} S_{2,1}
        \Biggr\}
        \nonumber\\ &&
        +\textcolor{blue}{C_A^2 N_F T_F} \Biggl\{
                -\frac{64 P_{11}}{(N-1) N^2 (N+1)^2 (N+2)^2}  S_{-2,1}
                -\frac{16 P_{20}}{9 (N-1) N^2 (N+1)^2 (N+2)^2} S_3
        \nonumber\\ &&
                -\frac{32 P_{21}}{3 (N-1) N^2 (N+1)^2 (N+2)^2} S_{-3} 
                -\frac{8 P_{22}}{9 (N-1) N^2 (N+1)^2 (N+2)^2} S_1^3 
        \nonumber\\ &&
                +\frac{16 P_{32}}{9 (N-1)^2 N^3 (N+1)^3 (N+2)^3} S_1^2 
                +\frac{16 P_{33}}{9 (N-1)^2 N^3 (N+1)^3 (N+2)^3} S_2 
        \nonumber\\ &&
                -\frac{8 P_{39}}{27 (N-1)^2 N^5 (N+1)^5 (N+2)^5} 
                +p_{qg}^{(0)}(N) \Biggl[
                        -\frac{32 P_{10}}{3 (N-1) N (N+1) (N+2)} S_{2,1}
        \nonumber\\ &&                
        +\Biggl(
                                -\frac{704}{3} S_3
                                +128 S_{2,1}
                                +512 S_{-2,1}
                        \Biggr) S_1
                        -512 S_{-3} S_1
                        -\frac{16}{3} S_1^4
                        -160 S_1^2 S_2
                        -16 S_2^2
                        -32 S_4
        \nonumber\\ &&
                        +\Biggl(
                                -192 S_1^2
                                +320 S_2
                        \Biggr) S_{-2}
                        -96 S_{-2}^2
                        +96 S_{-4}
                        -448 S_{2,-2}
                        -128 S_{3,1}
                        +512 S_{-3,1}
        \nonumber\\ &&
                        -768 S_{-2,1,1}
                        +192 S_1 \zeta_3
                \Biggr]
                +\frac{96 (N-2) (N+3) P_4}{(N-1) N^2 (N+1)^2 (N+2)^2}  \zeta_3
        \nonumber\\ &&
                +\Biggl(
                        \frac{8 P_{19}}{3 (N-1) N^2 (N+1)^2 (N+2)^2} S_2
                        -\frac{8 P_{36}}{27 (N-1)^2 N^4 (N+1)^4 (N+2)^4} 
                \Biggr) S_1
        \nonumber\\ &&
                +\Biggl(
                        -\frac{64 P_{13}}{(N-1) N^2 (N+1)^2 (N+2)^2} S_1
                        +\frac{32 P_{30}}{9 (N-1) N^3 (N+1)^3 (N+2)^3} 
                \Biggr) S_{-2}
        \Biggr\}
        \nonumber\\ &&
        + \textcolor{blue}{C_F^2 N_F T_F} \Biggl\{
                \frac{P_{31}}{N^5 (N+1)^5 (N+2)}
                -\frac{8 P_3}{3 N^2 (N+1)^2 (N+2)} S_1^3
                -\frac{16 P_6}{3 N^2 (N+1)^2 (N+2)} S_3 
        \nonumber\\ && 
                +\frac{64 P_{14}}{N^3 (N+1)^2 (N+2)} S_{-2} 
                -\frac{8 P_{23}}{N^3 (N+1)^3 (N+2)} S_1^2
                +\frac{8 P_{26}}{N^3 (N+1)^3 (N+2)} S_2
        \nonumber\\ &&
                + p_{qg}^{(0)}(N) \Biggl[
                        \Biggl(
                                -\frac{704}{3} S_3
                                +256 S_{2,1}
                        \Biggr) S_1
                        -256 S_{-3} S_1
                        -\frac{16}{3} S_1^4
                        -48 S_2^2
                        -160 S_4
                        -64 S_{-2}^2
        \nonumber\\ &&
                        -192 S_{-4}
                        -\frac{128}{N (N+1)} S_{2,1}
                        -128 S_{2,-2}
                        +64 S_{3,1}
                        +256 S_{-3,1}
                        -192 S_{2,1,1}
                \Biggr]
        \nonumber\\ &&
                +\frac{96 (N-1) \big(
                        3 N^2+3 N-2\big) }{N^2 (N+1)^2} \zeta_3
                - 256 \frac{2-N+N^2}{N^2 (N+1) (N+2)}
                        \left[S_{-2} S_1
                        - S_{-2,1}\right]
        \nonumber\\ &&
                +\Biggl(
                        -\frac{8 P_{28}
                        }{N^4 (N+1)^4 (N+2)}
                        -\frac{8 P_7 }{N^2 (N+1)^2 (N+2)}  S_2
                \Biggr) S_1
                -\frac{128 (N-1)}{(N+1)^2 (N+2)} S_{-3}
        \Biggr\}.
\label{eq:gqg2}
\end{eqnarray}
Here, $C_A = N_c, C_F = (N_c^2-1)/(2 N_c), T_F = 1/2$ and $N_c = 3$ in case of QCD, $\zeta_k, k \geq 2, k \in 
\mathbb{N}$ are the values of Riemann's $\zeta$ function,
$P_i$ denotes computed 
polynomials in $N$, $S_{\vec{a}}$ are the nested harmonic sums \cite{HSUM}
\begin{eqnarray}
S_{b,\vec{a}} \equiv S_{b,\vec{a}}(N) =  \sum_{k=1}^N \frac{({\rm sign}(b))^k}{k^{|b|}} S_{\vec{a}}(k),~~~S_\emptyset 
= 
1,~~b, a_i \in \mathbb{Z} \backslash \{0\},
\end{eqnarray}
and we used the shorthand notation 
\begin{eqnarray}
p_{qg}^{(0)}(N) = \frac{N^2 + N + 2}{N(N+1)(N+2)}~.
\end{eqnarray}
Despite of the emergence of non-first order factorizable terms in the higher order expansion terms in $\varepsilon$
in the master integrals, the method of Ref.~\cite{Ablinger:2017tan} allows to project on the terms which really
contribute to $O(1/\varepsilon)$. Due to this, we could calculate $\gamma_{qg}^{(2)}$ even in a massive
environment, without first encountering elliptic contributions which would only cancel in the very last step.
\section{The $\rho$-parameter at $O(a_s^3)$}
\label{sec:3}

\vspace*{1mm}
\noindent
The 3-loop QCD corrections to the $\rho$-parameter have been calculated in 
\cite{Grigo:2012ji}. There, some of the master integrals could not be calculated in 
closed form, and were determined from the beginning by a power series Ansatz in order to
finally derive numerical results. In Ref.~\cite{Ablinger:2017bjx}  we recently found 
an analytic solution of the equations of the type (\ref{eq:2}). Here the inhomogeneity 
$N(x)$ is always given in terms of harmonic polylogarithms with rational pre-factors.
The Wronskian $W(x)$ is given by a (factorizable) polynomial in $x$. For the 
homogeneous solutions one finds $_2F_1$ solutions \cite{IVH}. Let us consider, as an 
example, Eq.~(2.14) of \cite{Ablinger:2017bjx}. Its homogeneous solutions are given by
\begin{eqnarray}
\label{eq:ps1a}
\psi_{1a}^{(0)}(x) &=& \sqrt{2 \sqrt{3} \pi} 
\frac{x^2 (x^2-1)^2 (x^2-9)^2}{(x^2+3)^4}
\pFq{2}{1}{{\tfrac{4}{3}},\tfrac{5}{3}}{2}{z}
\\
\label{eq:ps2a}
\psi_{2a}^{(0)}(x) &=& \sqrt{2 \sqrt{3} \pi}
\frac{x^2 (x^2-1)^2 (x^2-9)^2}{(x^2+3)^4}
\pFq{2}{1}{{\tfrac{4}{3}},\tfrac{5}{3}}{2}{1-z},
\end{eqnarray}
with
\begin{eqnarray}
z \equiv z(x) = \frac{x^2(x^2-9)^2}{(x^2+3)^3}~.
\end{eqnarray}
This is not yet a solution in terms of complete elliptic integrals. Applying 
contiguous relations and the triangle group relations \cite{TAKEUCHI,Ablinger:2017bjx} one,
however, obtains
\begin{eqnarray}
\label{eq:ps1b}
\psi_{1b}^{(0)}(x) &=& \frac{\sqrt{\pi}}{4 \sqrt{6}} \Biggl\{
- (x-1)(x-3)(x+3)^2 \sqrt{\frac{x+1}{9-3x}}
\pFq{2}{1}{{\tfrac{1}{2}},\tfrac{1}{2}}{1}{z}
\nonumber\\
&& + (x^2+3)(x-3)^2     \sqrt{\frac{x+1}{9-3x}}
\pFq{2}{1}{{\tfrac{1}{2}},-\tfrac{1}{2}}{1}{z} \Biggr\}
\\
\label{eq:ps2b}
\psi_{2b}^{(0)}(x) &=&
\frac{2\sqrt{\pi}}{\sqrt{6}}\Biggl\{x^2 \sqrt{(x+1)(9-3x)}
\pFq{2}{1}{{\tfrac{1}{2}},\tfrac{1}{2}}{1}{1-z}
\nonumber\\ &&
+\frac{1}{8} \sqrt{(x+1)(9-3x)} (x-3)(x^2+3)
\pFq{2}{1}{{\tfrac{1}{2}},-\tfrac{1}{2}}{1}{1-z}\Biggr\},
\end{eqnarray}
with
\begin{eqnarray}
\label{eq:z1}
z  \equiv z(x) = -\frac{16x^3}{(x+1)(x-3)^3}~.
\end{eqnarray}
By using \cite{TRICOMI}
\begin{eqnarray}
\pFq{2}{1}{{\tfrac{1}{2}},\tfrac{1}{2}}{1}{z}  &=& \frac{2}{\pi} {\bf K}(z),~~~~~~~ 
\pFq{2}{1}{{\tfrac{1}{2}},-\tfrac{1}{2}}{1}{z} = \frac{2}{\pi} {\bf E}(z)
\end{eqnarray}  
one obtains the solution in terms of complete elliptic integrals of the first and 
second kind and may show that the elliptic integral of the second kind, {\bf E},
cannot be transformed away.  Since the $_2F_1$ functions have no representation as an 
integral in which the $z$ 
dependence appears in the integration bounds only, the inhomogeneous solution 
(\ref{eq:DI}) is an iterative non-iterative integral of the kind of 
Eq.~(\ref{eq:ITNEW}), \cite{Ablinger:2017bjx}.

\begin{figure}[t]
\centering
\includegraphics[width=0.47\textwidth]{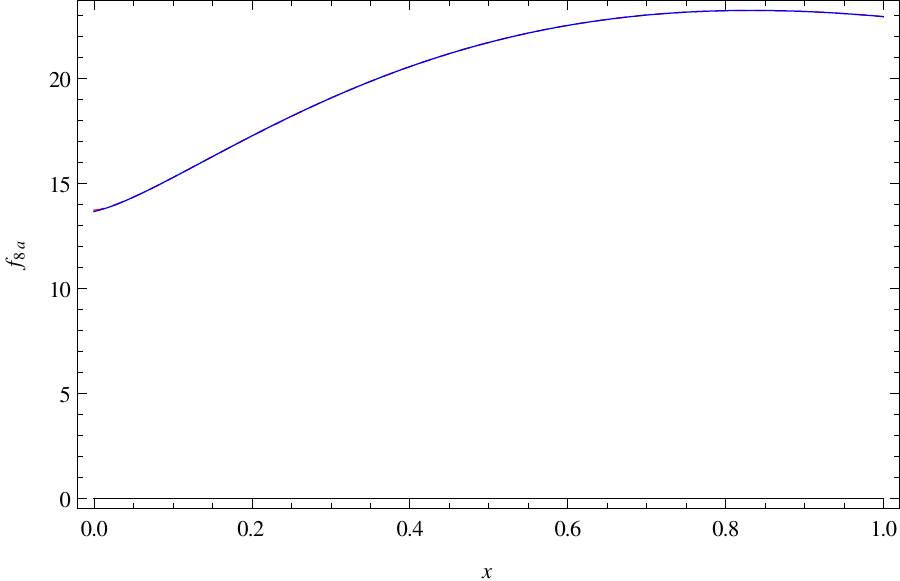}
\includegraphics[width=0.52\textwidth]{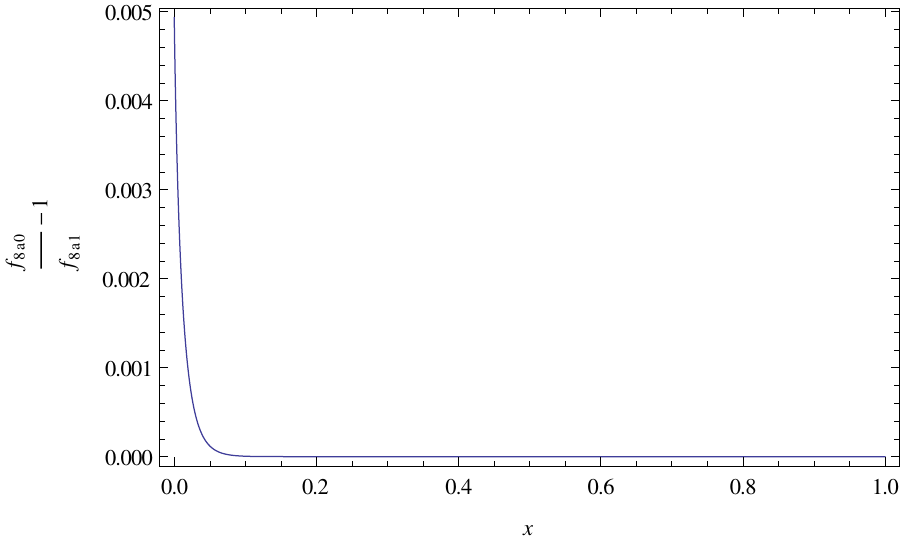}
\caption{ \small The inhomogeneous solution of Eq.~(2.14) of \cite{Ablinger:2017bjx}
as a function of $x$. Left panel: Red dashed line: expansion
around $x=0$; blue line: expansion around $x=1$. Right panel: illustration of the   
relative accuracy and overlap of the two solutions $f_{8a}(x)$ around 0 and 1; from 
Ref.~\cite{Ablinger:2017bjx}.
\label{fig:f8a}}
\end{figure}

One may now transform the inhomogeneous solution into power series solutions around 
$x=0$ and $x=1$ applying standard methods implemented in {\tt mathematica} and {\tt 
maple} to arbitrary order.  One obtains an accuracy of the overlapping solutions in the 
complete range $x \in [0,1]$, e.g. to an accuracy of $O(10^{-30})$ by the first fifty expansion 
terms. We show the solution in Figure~\ref{fig:f8a}.

We now would like to derive a $q$ series solution for the same problem. First the 
kinematic variable $x$ is expressed by $x = -1/y$, 
\begin{eqnarray}
z(x) \equiv k^2 =\frac{16 y}{(1-y)(1+3y)^3} = 
\frac{\vartheta^4_2(q)}{\vartheta^4_3(q)},
\label{eq:zx}
\end{eqnarray}  
with $\vartheta_k(q), k=1,2,3,4$ the Jacobi $\vartheta$-functions \cite{TRICOMI}.
One solves (\ref{eq:zx}) and obtains
\begin{eqnarray}
x = - \frac{\eta^4(2\tau) \eta^2(3\tau)}{\eta^2(\tau) \eta^4(6\tau)},
\label{eq:zxxx}
\end{eqnarray}  
which is a modular form multiplied by $1/\eta^{12}(\tau)$ and $\propto 1/q$. 

Some of the building blocks of the inhomogeneous solution are modular forms like
\begin{eqnarray}
\label{eq:Kmod}
{\bf K}(z) &=& \frac{\pi}{2}\sum_{k=1}^\infty \frac{q^k}{1+q^{2k}}
= \frac{\pi}{4} \overline{E}_{0;0}(i;1;q),~~~~~~\text{with} \\
\overline{E}_{n;m}(x; y; q) &=& \frac{1}{i} \left[
\ELI_{n;m}(x; y; q)
-\ELI_{n;m}(x^{-1}; y^{-1}; q)\right],~~n+m~~\text{even},
\end{eqnarray}  
while others are not. One example is {\bf E}$(z(x))$, given by
\begin{eqnarray}
{\bf E}(k^2)   &=& {\bf K}(k^2) + \frac{\pi^2 q}{{\bf K}(k^2)} \frac{d}{dq} 
\ln\left[\vartheta_4(q)\right].
\label{eq:EMain}
\end{eqnarray}
One has 
\begin{eqnarray}
q \frac{\vartheta_4'(q)}{\vartheta_4(q)} &=& - \frac{1}{2}\left[\ELI_{-1;0}(1;1;q) +
\ELI_{-1;0}(-1;1;q)\right]
+ \left[\ELI_{0;0}(1;q^{-1};q) + \ELI_{0;0}(-1;q^{-1};q)\right]
\nonumber\\ &&
- \left[\ELI_{-1;0}(1;q^{-1};q) + \ELI_{-1;0}(-1;q^{-1};q)\right].
\end{eqnarray}
Furthermore, $1/${\bf K}$(z(x))$ is given by
\begin{eqnarray}
\frac{1}{{\bf K}(k^2)} &=& \frac{2}{\pi \eta^{12}(\tau)}
\Biggl\{
\frac{5}{48}\Biggl\{
1 - 24 \ELI_{-1; 0}( 1; 1; q)
  - 4 \Biggl[
    1 - 24 \ELI_{-1; 0}( 1; 1; q^4)\Biggr]\Biggr\}
\nonumber\\ &&
\times
\Biggl\{
    -1 - 4 \Biggl[
    \ELI_{0; 0}( -1; 1/q; q^2) - 4 \ELI_{-1; 0}( -1; 1/q; q^2)
    + 4 \ELI_{-2; 0}( -1; 1/q; q^2)\Biggr]\Biggr\}
\nonumber\\ &&
- \frac{1}{16}\Biggl\{
  5 - 4 \Biggl[\ELI_{0; 0}( -1; 1/q; q^2) - 8 \ELI_{-1; 0}( -1; 1/q; q^2) +
       24 \ELI_{-2; 0}( -1; 1/q; q^2)
\nonumber\\ &&
- 32 \ELI_{-3; 0}( -1; 1/q; q^2) +
       16 \ELI_{-4; 0}( -1; 1/q; q^2)
       \Biggr]\Biggr\},
\label{eq:ONEoK}
\end{eqnarray}
which is a modular function. Terms of this kind do not allow to integrate the formal 
power series in $q$, since the coefficients of the  power series in $1/\eta(\tau)$ are not known in 
closed form. Therefore, the expressions obtained in the present case are of a more 
general nature than those appearing in \cite{BLOCH2,Adams:2014vja,Adams:2016xah}.
Another generalization appears due to the emergence of elliptic polylogarithm 
representations in which the parameters $x$ and $y$ become $q$-dependent.
\section{Conclusions}
\label{sec:4}

\vspace*{1mm}
\noindent
For single scale processes in the massless and massive cases to two-loop order one always has found 
iterative integral solutions over the alphabets given in \cite{Remiddi:1999ew,HSUM}
and the special numbers \cite{Blumlein:2009cf}. This applies also to the massless case
at 3-loop order \cite{Vogt:2004mw,Moch:2004pa,Vermaseren:2005qc}.
For the massive OMEs and Wilson coefficients in the asymptotic region 
$Q^2 \gg m^2$, with $Q^2$ the virtuality of the deep-inelastic process and $m$ the heavy quark mass, in the 
case of the iterative integral solutions \cite{Ablinger:2010ty,
Blumlein:2012vq,
Ablinger:2014vwa,
Ablinger:2014nga,
Ablinger:2014uka,
Ablinger:2014lka,
Behring:2014eya,
AGG,Ablinger:2015tua} 
more general alphabets contribute as well for the nested sums, iterated 
integrals, and in some cases also for the special constants  
\cite{Ablinger:2013cf,Ablinger:2011te, Ablinger:2014bra}. Beginning with 3-loop order 
one of the OMEs, $A_{Qg}^{(3)}$, \cite{Bierenbaum:2009mv,Ablinger:2017ptf} contains also
iterative non-iterative integrals caused by the emergence of the elliptic integrals of 
the first and second kind in some irreducible differential equations of second order.
These are directly related to the differential equations studied in 
\cite{Ablinger:2017bjx} for the 3-loop QCD corrections of the $\rho$-parameter.

We have shown that 3-loop anomalous dimensions can be computed in a massive 
environment in an automated way \cite{Ablinger:2017tan} from the $O(1/\varepsilon)$ 
term of the unrenormalized OME $\hat{\hat{A}}_{Qg}$. Using the method of arbitrary 
large moments \cite{Blumlein:2017dxp}, despite the fact that the master integrals contain 
elliptic terms at $O(\varepsilon^0)$ and higher, an assembly of all terms for 
fixed moments is possible, such that the final difference equations factorize at first 
order. A solution using  the packages {\tt Sigma, EvaluateMultiSums, SumProduction} and
{\tt HarmonicSums} is then possible.

This is likewise the case for 18 out of 28 color and $\zeta$-terms for the 
$O(\varepsilon^0)$ term as well,~cf.~\cite{Ablinger:2017ptf}. For the remaining color 
and $\zeta$-terms in part of the integrals the complete elliptic integrals of the 
first and second kind emerge, as well as integrals of other letters over them. The corresponding differential 
equations are related
to those emerging in the case of the 3-loop QCD corrections to the $\rho$-parameter 
by a variable transformation. We have outlined their solution in Section~\ref{sec:3}
leading to iterated non-iterative integral solutions. They possess analytic, fast converging 
series solutions around $x=0,1$, \cite{Ablinger:2017bjx}. We have also studied the associated $q$-series 
solutions. Here generalizations of previously studied elliptic solutions 
\cite{BLOCH2,Adams:2014vja,Adams:2016xah} occur. Due to the emergence of 
Dedekind-$\eta$ factors $\propto 1/\eta^k(\tau), k \in \mathbb{N}, k >0$, in front of modular forms expressed 
in (products of) Lambert-Eisenstein series, it is in general not possible to find
a diagonalization of the corresponding integrals, finally expressed in terms of elliptic polylogarithms.
However, elliptic polylogarithms can be widely used as a framework for the  
representation of the different building blocks emerging in the solution, also allowing their parameters $x$ and 
$y$, (\ref{eq:ELP}), to depend on $q$.


\end{document}